\documentclass[aps,twocolumn,showpacs,superscriptaddress,floatfix]{revtex4}

\usepackage[latin1]{inputenc}
\usepackage{graphicx}
\usepackage{amsmath, bbm,amssymb}

\begin{document}

\bibliographystyle{prsty}

\title{Phase-space theory for dispersive detectors of superconducting qubits}

\author{I. Serban}  

\affiliation{Physics Department, Arnold Sommerfeld Center for Theoretical Physics, and Center for NanoScience, \\
Ludwig-Maximilians-Universit\"at, Theresienstrasse 37, 80333 Munich, Germany} 

\affiliation{IQC and Dept.~of Physics and
Astronomy, University of Waterloo, 200 University Ave W, Waterloo, ON,
N2L 3G1, Canada}  

\author{E. Solano} 

\affiliation{Physics Department, Arnold Sommerfeld Center for Theoretical Physics, and Center for NanoScience, \\
Ludwig-Maximilans-Universit\"at, Theresienstrasse 37, 80333 Munich, Germany}  

\affiliation{Max-Planck Institute for
Quantum Optics, Hans-Kopfermann-Strasse 1, D-85748 Garching, Germany}

\affiliation{Secci\'{o}n F\'{\i}sica, Departamento de Ciencias,
Pontificia Universidad Cat\'{o}lica del Per\'{u}, Apartado 1761, Lima,
Peru} 

\author{F.K. Wilhelm}  

\affiliation{IQC and Dept.~of Physics and
Astronomy, University of Waterloo, 200 University Ave W, Waterloo, ON,
N2L 3G1, Canada}  

\date{\today}

\begin{abstract}
Motivated by recent experiments, we study the dynamics of a qubit quadratically coupled to its detector, a damped harmonic oscillator. We use a complex-environment approach, explicitly describing the dynamics of the qubit and the oscillator by means of their full Floquet state master
equations in phase-space.  We investigate the backaction of the environment on the measured qubit and explore several measurement protocols, which include a
long-term full read-out cycle as well as schemes based on short time transfer of information between qubit and oscillator. We also show that the pointer becomes measurable before all information in the qubit has been lost.
\end{abstract}


\maketitle

\section{Introduction}

The quantum measurement postulate is one of the most intriguing and
historically controversial pieces of quantum mechanics. It usually
appears as a separate postulate, as it introduces a
non-unitary time evolution. 

On the other hand, at least in principle, qubit and detector can be described by a coupled manybody Hamiltonian and thus the measurement process can be investigated using the
established tools of quantum mechanics of open systems. Even though
this does not lead to a solution of the fundamental measurement
paradox, such research gives insight into the {\em physics} of 
quantum measurement
 \cite{Peres93, Zurek81, Zurek93, Anderson01, Anderson01b, Adler03}.

This basic question has also gained practical relevance and has become 
a field of experimental physics in the context of quantum computing. 
Specifically, superconducting qubits have been proposed as building block of a scalable quantum computer \cite{Makhlin01,Devoret04,Nato06I,DiV06}. 
In these systems, the
detector is based on the same technology --- small, underdamped Josephson 
junctions --- as the device whose state is to be detected. Thus these circuits are an ideal test-bed to investigate the physics of 
quantum measurement. Implementing a measurement which is
fast and reliable, with a high (single-shot) resolution and high visibility is
a topic of central importance to the practical implementation of these
devices. 

The basic textbook version of a quantum measurement is based on von Neumann's 
postulate \cite{Cohen92, Neumann55}. The state of the system is projected onto 
the eigenstate of the observable being measured corresponding to
the eigenvalue being observed. This is not the only possible
quantum measurement and has been generalized to the idea of  
a positive operator-valued measure \cite{Nielsen00}. 

From the microscopic, Hamiltonian-based perspective, intensive research has been done on the measurement of small signals, which originated in the theory of gravitational wave 
detection \cite{Braginsky95}. The main challenge has been to identify how
signals below the limitations of the uncertainty relation of the detector can 
be measured - a regime in which the detector response is also strictly linear. 
This work has resulted in the notion of a quantum
nondemolition (QND) measurement~\cite{Braginsky95}, which is the closest to a microscopic formulation of
a von Neumann measurement. This result has been generalized to many other
systems, prominently atomic physics, and also found its way to the superconducting qubits literature. Here, the analogy of a tiny signal is the limit
of weak coupling between qubit and detector. Another body of work 
\cite{Zurek81,Zurek93} takes a more general starting point and discusses the 
relevance of pointer state and environment induced superselection. 

The measurement techniques used in superconducting qubits are covering many of
the mentioned situations. Weak measurements can be performed using single
electron transistors. Based on weak measurement theory, this is well
understood (see Ref.~\cite{Makhlin01} and the references therein) 
but only of limited use for
superconducting qubits. There the measurements are far from projective, their resolution is in practice rather limited and the whole process is very slow. 
In the case of qubit, the task is {\em not}  to amplify an arbitrarily weak signals, but to discriminate two states in the best possible way. If the detector can be decoupled from the qubit when no measurement needs to be performed, this discrimination may involve strong qubit-detector coupling \cite{Scripta02}. 

An opposite, generic approach is to perform a switching measurement --- the detector switches out of a metastable state depending on the state of qubit. Switching is a highly nonlinear phenomenon, so this type of detection is far from the weak measurement scenario. In most of the early generic setup, this  process is a switching of a superconducting device, e.g. a superconducting quantum interference device (SQUID), from the superconducting to the 
dissipative state \cite{Science00,EPJB03,Vion02,Martinis02}. 
This technique goes a long way, and some experiments have proven
that the switching type of readout can achieve high contrast \cite{I.Chiorescu03212003, steffen:050502}. It has the drawback that it is not a projective readout 
and during the switching process hot quasiparticles with a long relaxation time are created. This limits the time between the consecutive measurements. Parts of this technique are well understood, such as the
switching histogram \cite{Martinis87}, the pre-measurement backaction \cite{PRBR033},
and the influence of the shunt impedance to the SQUID \cite{Joyez99,PRB051}, but
there is no full and single theory of this process on the same level of detail as 
the weak measurement theory. 

Recent developments of detection schemes have lead to vast improvements based
on two innovations: instead of directly measuring a certain observable 
pertaining to a qubit state, one uses a pointer system, and measures one of its observables influenced by the state of the qubit. The 
observation is usually materialized in the frequency shift of an appropriate
resonator, whose response to an external excitation links to the
measurement outcome \cite{Lupascu04,Lee05,Blais04,Schuster05}. These measurements offer good sensitivity, high visibility \cite{Wallraff05}, and fast repetition rates. They also allow to keep the 
qubit at a well-defined operation point, although not always the optimum one. 
 In many cases, the resonators in use
are nonlinear --- based on Josephson junctions. Thus, at stronger excitation,
generic nonlinear effects can be exploited. These nonlinear effects go 
up to switching, which in contrast to the critical current switching 
is between two dissipationless states \cite{Siddiqi05}. 
Due to this performance and 
versatility, these devices also offer an ideal example for investigating the
crossover between weak and strong measurements and the role of nonlinearity.

Analyzing the properties of quantum measurement is an application of open 
quantum systems theory: the backaction contains a variant of projection which can be viewed in an ensemble as dephasing. The resolution is determined by the
behavior of the detector under the influence of the qubit viewed as an 
environment. For open quantum systems, a number of tools have been developed. 
Most of them, prominently Born and Born-Markov master equations (see e.g.~
\cite{Nato06II} and references therein for a recent review) assume weak 
coupling between qubit and environment and are hence a priori unsuitable for 
studying strong qubit-detector coupling. Tools for stronger coupling have
been developed \cite{Weiss99,Keil01} but are largely restricted to harmonic oscillator baths and hence unsuitable to treat the generally nonlinear physics of the systems of interest. 
The Lindblad equation \cite{Lindblad76} is claimed to be valid up to strong coupling, however, due to its strong Markovian assumption it is 
unsuitable for strongly coupled superconducting systems. 

In this paper we present a theoretical tool allowing to 
describe dispersive measurements involving nonlinearities. The tool is developed
alongside the example of the experimental setup studied in Ref.~\cite{Lupascu04}. It is based on the
complex environments approach similar to what is used in cavity QED 
\cite{Walls94} but also in condensed-matter open quantum systems 
\cite{Paladino02,Thorwart03,Dykman87}. The idea is to introduce the potentially strongly and
nonlinearly coupled component of the detector as part of the quantum system and
only treat the weakly coupled part as an environment. In other words, we single out one prominent degree of freedom of the detector from the rest and treat it on equal footing with the qubit. This "special treatment" of one environmental degree of freedom is an essential point of this approach because it allows us to describe the dynamics of a qubit coupled arbitrarily strong to a non-Markovian environment. On the other hand it enhances the dimension of the Hilbert space to be captured. This technical  complication can be handled
using a phase space representation of the extra degree of freedom --- in our example, a harmonic oscillator. 

In Section \ref{section_model} we derive the model Hamiltonian motivated by Ref.~\cite{Lupascu04}. For this Hamiltonian we derive in Section \ref{section_method} a master equation and present  a phase space method which enables us to analyze the dynamics of an infinite level system. In Section \ref{section_results} we demonstrate that our method enables to extract informations about the measurement process of the qubit, such as dephasing and measurement time and also present three different measurement protocols. 

\section{From circuit to Hamiltonian}  \label{section_model}

\begin{figure}[!h]
  \includegraphics[width=.4\textwidth]{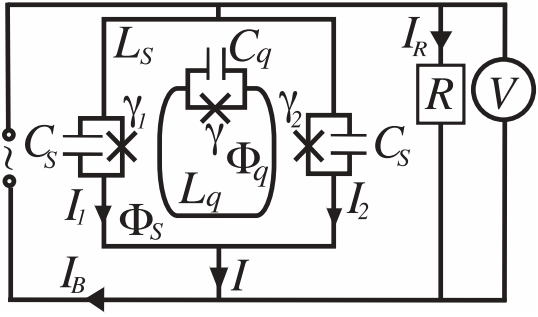} 
 \caption{Simplified circuit consisting of a qubit with one Josephson
junction (phase $\gamma$, capacitance $C_q$ and inductance $L_q$)
inductively coupled to a SQUID with two identical junctions (phases
$\gamma_{1,2}$, capacitance $C_S$) and inductance $L_S$. The SQUID is
driven by an ac bias $I_B(t)$ and the voltage drop is measured
by a voltmeter with internal resistance $R$. The total flux through
the qubit loop is $\Phi_q$ and through the SQUID is
$\Phi_S$ .}  \label{circuit} 
\end{figure}

We consider a simplified
version of the experiment described in Ref.~\cite{Lupascu04}.  The
circuit consists of a flux qubit drawn in the single junction version,
the surrounding SQUID loop, an ac source, and  a shunt
resistor, as depicted in Fig.~\ref{circuit}. We note here that we later approximate the qubit as a two-level system. The qubit used in the actual experiment contains three junctions. An analogous but less transparent derivation would, after performing the two-state approximation, lead to the same model, parameterized by the two-state Hamiltonian, the circulating current, and the mutual inductance, in an identical way \cite{EPJB03}.
The measurement process is started by switching on the ac 
source and monitoring the amplitude and/or phase of the voltage drop
across the resistor. 

The  SQUID acts as an oscillator whose resonance frequency depends on
the state of the qubit. When the measurement is started, the qubit
entangles with the resonator an shifts its frequency.  The value of
the oscillator frequency relative to the frequency of the ac driving
current determines the amplitude and phase of the voltage drop across
the resistor.

The detector (voltmeter) contains an internal resistor. This is a
dissipative element connecting the quantum mechanical system (qubit +
SQUID) to the macroscopic observer.  The resistor is needed for
performing the measurement, defined as the transfer of quantum
information encoded in a superposition of states to classical
information encoded in the probabilities with which the the voltmeter
shows certain values, e.g.~voltage amplitude or phase.   Note that practically this resistor may be the distributed 
impedance of the coaxial line connected to the chip.

In this section we derive an effective Hamiltonian for
this system. Our starting point is a set of current conservation
equations for the circuit of Fig.~\ref{circuit}.

The total magnetic fluxes through the SQUID and the qubit loops can be
divided into
screening $(s)$ fluxes produced through circulating currents and
external $(x)$ fluxes from outside sources $
\Phi_S=\Phi_{S}^{(x)}+\Phi_{S}^{(s)}$ and
$\Phi_q=\Phi_{q}^{(x)}+\Phi_{q}^{(s)}$. Generally, Ampere's law for a
system of current loops can be represented in matrix form
\begin{eqnarray}  \left(\begin{matrix} \Phi_{q}^{(s)}   \\
\Phi_{S}^{(s)} \end{matrix}\right)&=&- \left(\begin{matrix} L_q &
M_{Sq}\\ M_{Sq}& L_S \end{matrix}\right) \left(\begin{matrix} I_q \\
I_S \end{matrix}\right),
\end{eqnarray}  which can be inverted
\begin{eqnarray}  \left(\begin{matrix}I_q\\ I_S\end{matrix}\right)&=&
-\frac{1}{M_{\Sigma}^2}\left(\begin{matrix}L_S & -M_{Sq}\\-M_{Sq}&
L_q\end{matrix}\right) \left(\begin{matrix}\Phi_{q}^{(s)}\\
\Phi_{S}^{(s)}\end{matrix}\right)\label{matrix_currents} ,
\end{eqnarray}  where $M_{Sq}$ is the mutual inductance and $
M_{\Sigma}$ is the determinant of the inductance matrix  $
M_{\Sigma}^2=L_qL_S-M_{Sq}^2> 0$. The circulating current through the
SQUID loop is given by the difference of the currents through the two
branches $I_S=(I_1-I_2)/2$ and the bias current is $I=I_1+I_2$.  The
fluxoid quantization  \cite {Tinkham96} in the two loops reads
\begin{eqnarray}
\Gamma_-&=&\gamma_1-\gamma_2=2\pi\frac{\Phi_S}{\Phi_0}{\rm
mod}\label{flux1} 2\pi,\\ \gamma&=&2\pi\frac{\Phi_q}{\Phi_0}{\rm mod}
2\pi,\label{flux2}
\end{eqnarray}   where $\Phi_0=h/2e$ is the magnetic flux quantum for a 
superconductor.  To obtain
the equations of motion for the phases $\gamma, \Gamma_{\pm}$ with
$\Gamma_{\pm}=\gamma_1\pm \gamma_2$ we start from the current conservation in
each node. 
\begin{eqnarray} I_j&=&I_{cS}\sin\gamma_j+\dot{V}C_S\nonumber\\
&=&I_{cS}\sin\gamma_j+\ddot{\gamma}_j\frac{\Phi_0}{2\pi}C_S,\label{branch12}\\
j&\in&\{1,2\}\nonumber
\end{eqnarray}  Here we assume that the two junctions have identical
critical currents. This symmetry, as will be discussed below, has the
consequence that at zero bias current through the SQUID, the qubit
will be isolated from its environment. In experiment, the two SQUID
junction will of course not be identical. For an asymmetric SQUID the
qubit can be protected from environmental noise \cite{Bertet05b} by
applying an appropriate DC bias. 

Using Eqs.~(\ref{matrix_currents}) and (\ref{branch12}) for the circulating current, we obtain
\begin{eqnarray}
I_{cS}\cos\frac{\Gamma_+}{2}\sin\frac{\Gamma_-}{2}&+&\ddot{\Gamma}_-\frac{\hbar}{4e}C_S
\nonumber\\ &=&
\frac{1}{M_{\Sigma}^2}\left(M_{Sq}\Phi_{q}^{(s)}-L_q\Phi_{S}^{(s)}\right)\label{circS}.
 \end{eqnarray} Considering the analogy between Josephson junctions
and inductors we introduce the Josephson inductance
$L_{JS}=\Phi_0/(2\pi I_{cS})$ and rewrite Eq.~(\ref{circS})
using the fluxoid quantization (\ref{flux1}) and (\ref{flux2})
\begin{eqnarray}
\!\!\!\!\!\!\! \frac{1}{L_{JS}}\cos\frac{\Gamma_+}{2}\sin\frac{\Gamma_-}{2}&+&\ddot{\Gamma}_-\frac{C_S}{2}\nonumber\\
&=&\frac{1}{M_{\Sigma}^2}\left(M_{Sq}\gamma-L_q\Gamma_-\right)+\Xi_1 , \label{circS2}
\end{eqnarray} 
where the influence of external fields is captured in
$\Xi_1=2\pi/(\Phi_0M_{\Sigma}^2)\left(-M_{Sq}\Phi_{q}^{(x)}+L_q\Phi_{S}^{(x)}\right)$.                          

For the bias current we have $I+V/R=I_B(t)$ and from
Eq.~(\ref{branch12}) we obtain
\begin{eqnarray}
\frac{1}{L_{JS}}\sin\frac{\Gamma_+}{2}\cos\frac{\Gamma_-}{2}+\ddot{\Gamma}_+\frac{C_S}{2}+\frac{1}{4R}\dot{\Gamma}_+&=&\frac{\pi}{\Phi_0}I_B(t) . \nonumber \\ \label{bias} 
\end{eqnarray} 
 
For the circulating current in the qubit loop it follows from
Eq.~(\ref{matrix_currents})
\begin{eqnarray}  I_q&=&C_q\ddot{\gamma}\frac{\hbar}{2e}+
I_{cq}\sin\gamma\nonumber\\ &=&
-\frac{1}{M_{\Sigma}^2}\left(L_s\Phi_{q}^{(s)}-M_{Sq}\Phi_{S}^{(s)}\right)
\end{eqnarray}  
Using $L_{Jq}=\Phi_0/(I_{cq}2\pi)$,
Eqs. (\ref{flux1}), and (\ref{flux2}), this becomes 
\begin{eqnarray} C_q\ddot{\gamma}+ \frac{1}{L_{Jq}}\sin\gamma&=&
-\frac{L_S}{M_{\Sigma}^2}\gamma+
\frac{M_{Sq}}{M_{\Sigma}^2}\Gamma_-+\Xi_2 \label{circQ},
\end{eqnarray} where we defined
$\Xi_2=2\pi/(M_{\Sigma}^2\Phi_0)\left(-M_{Sq}\Phi_{S}^{(x)}+L_S\Phi_{q}^{(x)}\right)$. 

From Eqs.~(\ref{circS2}), (\ref{bias}), and (\ref{circQ}), we observe that
$\Gamma_+$, the phase drop across the SQUID, serves as a pointer: it
couples to the qubit degree of freedom $\gamma$ and is read out by the
classical observer, which appears in the classical equation of motion
(\ref{bias}) as a dissipative term.  Without bias current $I_B=0$, the
classical solution for this degree of freedom becomes $\Gamma_+=0$
independent of the internal degree of freedom $\Gamma_-$ and the
qubit.  It follows that, in the absence of $I_B$, there is no coupling
between the quantum mechanical system and the environment, as the
pointer is decoupled from the observer. 

 We start the derivation of the system Hamiltonian suppressing the
dissipative term in Eq. ~(\ref{bias}). It will be later reintroduced
in the form of an oscillator bath.  Starting from the equations of
motion (\ref{circS2}), (\ref{bias}), and (\ref{circQ}), for $\Gamma_{\pm}$ and
$\gamma$ we first determine the Lagrangian such that $
d_t(\partial_{\dot{\gamma}}\mathcal{L})=\partial_{\gamma}\mathcal{L}$  and $
d_t(\partial_{\dot{\Gamma}_\pm}\mathcal{L})=\partial_{\Gamma_\pm}\mathcal{L}$.  We introduce the canonically conjugate momenta
$p=\partial_{\dot{\gamma}}\mathcal{L}=\hbar^2C_q\dot{\gamma}/e^2$
and $P_{\pm}=\partial_{\dot{\Gamma}_\pm}\mathcal{L}=\hbar^2C_s\dot{\Gamma}_\pm/(2e^2)$ and finally derive
the Hamiltonian using
$\mathcal{H}=\dot{\gamma}p+\dot{\Gamma}_-P_-+\dot{\Gamma}_+P_+-\mathcal{L}$. This
leads to
\begin{eqnarray}
\mathcal{H}& \!\! = \!\! & \left(\frac{P_+^2+P_-^2}{C_s}
+\frac{p^2}{2C_q}\right)\frac{e^2}{\hbar^2} -
\Bigg(\frac{2}{L_{JS}}\cos\frac{\Gamma_-}{2}\cos\frac{\Gamma_+}{2}\nonumber\\
&+&\frac{M_{Sq}}{M_{\Sigma}^2} \gamma\Gamma_--
\frac{L_{q}}{M_{\Sigma}^2}\frac{\Gamma_-^2}{2}+\Xi_1\Gamma_-\nonumber\\
&+&\frac{1}{L_{Jq}}\cos{\gamma}-
\frac{L_{S}}{M_{\Sigma}^2}\frac{\gamma^2}{2}+\Xi_2\gamma+\frac{e}{\hbar}I_B(t)\Gamma_+
\Bigg)\frac{\hbar^2}{e^2} .
\end{eqnarray} 
Now we proceed to simplify this Hamiltonian using the
assumptions that $L_{JS}\gg L_S$, which
applies to small SQUIDs as the ones used for qubit readout, and that
the driving strength is small enough to remain in the harmonic part of
the potential $|I_B|\ll I_{cS}$. 
Using the latter assumption, we can
expand the potential energies to second order around the minimum and
obtain two  coupled harmonic oscillators ($\Gamma_+$ and $\Gamma_-$)
with greatly different frequencies. $\Gamma_-$ evolves in a much
narrower potential ($\propto1/L_S$) than that of $\Gamma_+$ ($\propto
1/L_{JS}$). Therefore we can perform an adiabatic approximation and
substitute $\Gamma_-$ through its average position. We obtain the
following potential for the remaining degree of freedom $\Gamma_+$
\begin{eqnarray}
U&=&U_0+\frac{1}{4L_{JS}}\cos\left(\Xi_1\frac{M_{\Sigma}^2}{2L_q}+\gamma\frac{M_{Sq}}{2L_q}\right)\nonumber\\
&\times&\left(\Gamma_+-\frac{I_B(t)}{I_{cS}}
\frac{1}{\cos\left(\Xi_1\frac{M_{\Sigma}^2}{2L_q}+\gamma\frac{M_{Sq}}{2L_q}\right)}\right)^2\frac{\hbar^2}{e^2}.
\end{eqnarray}  
In the next step, we perform the two-state
approximation of the qubit along the lines of
Ref.~\cite{Mooij99,Weiss99}, reducing its dynamics to the two lowest
energy eigenstates. This space is spanned by wave functions  centered
around two values $\hat{\gamma}=\gamma_0\hat{\sigma}_z$  ($\gamma$
either in the left or the right well of the potential). 
While the manipulation of the qubit is usually performed at
the optimum working point \cite{Vion02}, the readout can and should
be performed in quantum nondemolition (QND) measurement i.e.~in the pure
dephasing limit. This reduces the qubit Hamiltonian to
$\epsilon_0\hat{\sigma}_z$.  We allow
for a significant off-diagonal term $\propto\hat{\sigma}_x$ to have acted in the 
past in order to prepare superpositions of eigenstates of $\sigma_z$. Physically,
this situation is achieved by either making one of the qubit junctions tunable,
or imposing a huge energy bias to the qubit. 

We note here that if, opposed to the case we will discuss in the following, the measurement interaction would not commute with the qubit Hamiltonian, a full analysis in terms of quantum measurement theory would be required. Similar to Refs.~\cite{Korotkov99, Ruskov02} the action on the system given by each measurement result would need to be determined in order to quantify the information that the observer can obtain about the initial state of the qubit, as well as the state following the measurement.

After these approximations the qubit-SQUID Hamiltonian reads
\begin{eqnarray}
\!\!\!\!\!\!\!\! \hat{H}_S& \!\! = \!\! &\epsilon(t)\hat{\sigma}_z+\frac{\hat{P}_+^2}{2m}+\frac{m(\Omega^2+\Delta^2\hat{\sigma}_z)}{2}\hat{x}^2-F(t)\hat{x}  \, , \label{sysH}
\end{eqnarray}  where $\hat{x}$  corresponds to the external degree of
freedom of the SQUID $\Gamma_+$ and $F(t)=F_0\sin(\nu t)$ originates
in the ac driving by a classical field. 
The conversion of the parameters to circuit-related
quantities can be found in Appendix \ref{Apcirc}.  Here $\Delta$ is the
quadratic frequency shift (QFS).

An important property of this Hamiltonian is the absence of the commonly used linear
coupling between the two-level system and the harmonic oscillator~\cite{EPJB03}. In our case the qubit couples to the
squared coordinate of the oscillator, which leads to a qubit dependent
change in the frequency of the harmonic oscillator instead of the
shift of the potential minimum.   

Because of the coupling to the
driven oscillator the qubit energy splitting becomes time-dependent
$\epsilon(t)=\epsilon_0+\upsilon I_B^2(t)$.

To model the dissipation introduced by the resistor we
follow the standard Caldeira-Leggett approach
\cite{Caldeira81,Leggett84,Leggett87,Weiss99} and include an
oscillator bath to our Hamiltonian
\begin{eqnarray}
\hat{H}&=&\hat{H}_S+\underbrace{\sum_i\left(\frac{\hat{p}_i^2}{2m_i}+\frac{m_i\omega_i^2}{2}\hat{y}_i^2\right)}_{\hat{H}_B}+\underbrace{\hat{x}\sum_i\lambda_i
\hat{y}_i}_{\hat{H}_{SB}}\label{bathH}.
\end{eqnarray}  with
$J(\omega)=\sum_i\lambda_i^2\frac{\hbar}{2m_i\omega_i}\delta(\omega-\omega_i)=m\hbar\kappa\omega
\Theta(\omega-\omega_c)/\pi$ \cite{Ingold98} where $\Theta$ is the
Heaviside step function and $\kappa=[s^{-1}]$ the photon loss
rate. The cut-off frequency $\omega_c$ is physically motivated by the
high frequency filter introduced by the capacitors.  

\section{Method}\label{section_method}
Our goal is to analyze the resolution and measurement time and investigate the 
backaction on the qubit. The former requires tracing over the qubit and 
discuss the dynamics of the pointer variable of the detector, the latter
requires tracing over the detector degrees of freedom. 

It is well established how to do this in principle exactly \cite{Weiss99}
when the qubit couples to a Gaussian variable of the detector (i.e.~sum of quadratures of the environmental coordinates). A method to map a damped harmonic oscillator to bath of uncoupled oscillators with a modified spectral density  \cite{Garg85,Ambegaokar05} also exists. In our case, due to the quadratic coupling between the qubit and the damped oscillator (\ref{sysH}), any such normal-mode
transform does not lead to the usual Gaussian model and thus 
cannot use many of the methods developed for the spin-boson model.  

There are several approaches to dealing with this challenge. As long as the  coupling is weak, $\Delta\ll\Omega$, one can still linearize the detector dynamics and make a Gaussian approximation as it was done in Ref.~\cite{Blais04}. Nevertheless, weak coupling decoherence theory as reviewed e.g.~in Refs.~\cite{Makhlin01,Nato06II} builds on two-point correlators and cannot distinguish Gaussian from non-Gaussian environments. 

In this work we describe arbitrarily large couplings between qubit and 
oscillator going beyond the Gaussian approximation. The only small parameter we rely upon is the decay rate of the oscillator $\kappa$. This is justified by the fact that dispersive measurement only makes sense for large oscillator
quality factors $Q>1$. We treat a composite quantum system --- qubit $\otimes$ oscillator --- weakly coupled to the heat bath represented by
the resistor. This {\em complex environments approach} resembles the methods
of, e.g.~Refs.~\cite{Paladino02,Blais04}. 

We start with the standard master equation for the reduced density operator in Schr\"odinger picture and Born-Markov approximation \cite{Nato06II, Blum96}, 
assuming factorized initial conditions $\rho(0)=\rho_S(0)\otimes \rho_B(0)$
\begin{eqnarray}
&&\frac{d}{dt}\hat{\rho}_S(t)=
\frac{1}{\mathbbm{i}\hbar}[\hat{H}_S,\hat{\rho}_S(t)]\label{mastereq} \\ 
&+&
\frac{1}{(\mathbbm{i}\hbar)^2}\int_0^t dt'\:{\rm Tr}_B \: \left[\hat{H}_{SB},[\hat{H}_{SB}(t,t'),\hat{\rho}_S(t)\otimes\hat{\rho}_B(0)]\right],\nonumber
\end{eqnarray}
where  $\hat{H}_{SB}(t,t')={ \hat{U}_{t'}^{t} }\hat{H}_{SB}\hat{U}_{t}^{t'}$  and $\hat{U}_t^{t'}=\mathcal{T}\exp\left(\int_t^{t'}d\tau(\hat{H}_S+\hat{H}_B)/(\mathbbm{i}\hbar)\right)$ and $\mathcal{T}$ is the time-ordering operator. 
In thermal equilibrium
there will be correlations between the main oscillator and the
oscillators of the bath, so the initial state is not strictly speaking factorized. In the low $\kappa$ limit here, these correlations
will affect the dynamics only in higher order $\kappa^2$ and can
hence be neglected. This is a standard assumption in the perturbative 
treatment of open systems where here $\kappa$ is the perturbative 
parameter, see e.g. \cite{Nato06II}
This approach is valid at finite temperatures $k_BT \gg  \hbar \kappa$, for times $t\gg1/\omega_c$ \cite{Alicki06,Nato06II}, which is
the limit we will discuss henceforth. We assume unbiased noise 
$ \langle \hat{H}_{SB}\rangle=0$. 
The Hamiltonian (\ref{sysH}) describes a driven harmonic oscillator, therefore the 
Floquet modes (see e.g.~Refs.~\cite{Hanggi98,Grifoni98} for a short review) form the appropriate basis in which we express the master equation. 
For a driven harmonic oscillator the Floquet modes \cite{Grifoni98} are given by
 \begin{eqnarray}
&&\Psi_{n}(x,t)=\varphi_{n}(x-\xi(t))\nonumber\\
&\times&\exp\left[\frac{i}{\hbar}\left(m\dot{\xi}(t)(x-\xi(t))-E_{n}t+\int_0^tdt'L(t')\right)\right]\nonumber,\\
&=&\Phi_n(x,t)\mathbbm{e}^{-\mathbbm{i}E_n t/\hbar},
\end{eqnarray}
where $E_n=\hbar\Omega(n+1/2)$, $\varphi_n(x)$ a number state, $\xi(t)$ is the classical trajectory and $L(\xi,\dot{\xi},t)$ the classical Lagrangian of the driven undamped  oscillator 
\begin{eqnarray}
\xi(t)&=&\frac{F_0\sin(\nu t)}{m(\Omega^2-\nu^2)},\\
L(\xi,\dot{\xi},t)&=&\frac{1}{2}m\dot{\xi}^2(t)-\frac{1}{2}m\Omega^2\xi^2(t)+\xi(t)F(t).
\end{eqnarray}
We also define the operator $\hat{A}$ as the annihilation operator corresponding to a Floquet mode
\begin{eqnarray}
\hat{a}&=&\hat{A}+\zeta(t),
\end{eqnarray}
where $\zeta(t)=\sqrt{\frac{m}{2\hbar\Omega}}(i\dot{\xi}(t)+\Omega\xi(t))$ so that $\hat{A}\Phi_n(x,t)=\sqrt{n}\Phi_{n-1}(x,t)$. 
After some algebra we obtain 
\begin{eqnarray}
\hat{x}&=&\sqrt{\frac{\hbar}{2m\Omega}}(\hat{A}+\hat{A}^{\dagger})+\xi(t),\\
\hat{a}(t,t')&=&\mathbbm{e}^{\mathbbm{i}\Omega(t-t')}\hat{A}+\zeta(t').\label{att}
\end{eqnarray}
Eq.~(\ref{att}) has been obtained by calculating  $\hat{a}(t,t')\Phi_n(x,t)$, where $\{\Phi_n(x,t)\}$ build a complete set of functions at any time $t$. Here one can interpret the sum $\hat{A}+\hat{A}^{\dagger}$ as the deviation of $\hat{x}$ from the classical trajectory $\xi(t)$. 

Since we are describing a composite quantum mechanical system, the operators in Eq.~(\ref{mastereq}) can be written in the qubit $\hat{\sigma}_z$ basis as follows
\begin{eqnarray}
\!\!\!\!\!\!\!\! \hat{\rho}_S & \!\! = \!\! & \left(\begin{matrix}
  \hat{\rho}_{\uparrow\uparrow}&\hat{\rho}_{\uparrow\downarrow} \\ \hat{\rho}_{\downarrow\uparrow}&\hat{\rho}_{\downarrow\downarrow}\end{matrix}\right),\:
  \hat{H}_S=\left(\begin{matrix}
\hat{H}_{S\uparrow}&0\\0&\hat{H}_{S\downarrow}
\end{matrix}\right)\label{densityqubit} \\
\!\!\!\!\! \hat{A}& \! = \! &\left(\begin{matrix}
\hat{A}_{\uparrow}&0\\0&\hat{A}_{\downarrow}
\end{matrix}\right),\:\:
\hat{a}(t,t')=\left(\begin{matrix}
\hat{a}_{\uparrow}(t,t')&0 \\ 0&\hat{a}_{\downarrow}(t,t')
\end{matrix}\right) ,\\
\sigma&\in&\{\uparrow,\downarrow\},\nonumber
\end{eqnarray}
where all the matrix elements are operators in the oscillator Hilbert space
\begin{eqnarray}
&& \!\!\!\!\!\!\!\!\!\!\!\!\!\!\! \hat{H}_{S\uparrow,\downarrow}  \! = \! \pm\epsilon(t) \! + \! \left(\frac{\hat{P}_+^2}{2m} +\frac{m(\Omega^2\pm\Delta^2)}{2}\hat{x}^2 - \hat{x}F(t) \! \right) , \\
&& \hat{a}_{\sigma}(t,t')  =  \hat{A}_{\sigma}\mathbbm{e}^{\mathbbm{i}\Omega_{\sigma}(t-t')}+\zeta_{\sigma}(t')
\end{eqnarray}
and $\hat{A}_{\uparrow,\downarrow}$ is the annihilation operator of a Floquet mode with frequency $\Omega_{\uparrow,\downarrow}=\sqrt{\Omega^2\pm\Delta^2}$. The functions $\zeta(t)$ and $\xi(t)$ also depend on the frequency of the harmonic oscillator, therefore they become $2\times2$ diagonal matrices. 

As we observed in the previous section, as long as $I_B=0$ there is no direct coupling between the qubit and the oscillator in the second order approximation and thus no coupling to the environment. Therefore, at $t=0$, before one turns on the ac driving, the harmonic oscillator has the frequency $\Omega$ independent of the qubit. Therefore the initial condition for the density matrix $\hat{\rho}_S$ is $\hat{\rho}_S(0)=\hat{\rho}_{\rm qubit}\otimes\hat{\rho}_{HO}^{(\Omega)}$. 

We introduce also the annihilation operator $\hat{A}_0$ for the Floquet modes with frequency $\Omega$ which relates to $\hat{A}_{\sigma}$ as follows
\begin{eqnarray}
\hat{A}_{\sigma}&=&
          \frac{1}{2}\left( \hat{A}_0^{\dagger}\left( f_{\sigma}-f_{\sigma}^{-1}\right)+\hat{A}_0\left(f_{\sigma}+f_{\sigma}^{-1}\right)\right)\nonumber\\
             &-&\zeta_{\sigma}^*(t)+{\rm Re}\zeta_0(t)f_{\sigma}-\mathbbm{i}{\rm Im}\zeta_0(t)\frac{1}{f_{\sigma}}, \label{creation0}
\end{eqnarray}
where $f_{\sigma}=\sqrt{\Omega_{\sigma}/\Omega}$. 

Using the operators introduced above in Eq.~(\ref{mastereq}), we obtain 
\begin{widetext}
\begin{eqnarray}
\dot{\hat{\rho}}_{\sigma\sigma'}(t)&=&
   \frac{1}{\mathbbm{i}\hbar}\hat{H}_{S\sigma}\hat{\rho}_{\sigma\sigma'}(t)
  -\frac{1}{\mathbbm{i}\hbar}\hat{\rho}_{\sigma\sigma'}(t)\hat{H}_{S\sigma'}
  +\frac{1}{(\mathbbm{i}\hbar)^2}\int_0^t\!\!dt'\int_0^{\infty}\!\!\!d\omega J(\omega)\nonumber\Big\{\!
\left(\mathbbm{e}^{\mathbbm{i}\omega (t-t')}n(\omega)+\mathbbm{e}^{-\mathbbm{i}\omega (t-t')}(n(\omega)+1)\right)\nonumber\\
   &&\left(g_{\sigma}^2\left(\hat{a}_{\sigma}+\hat{a}_{\sigma}^{\dagger}\right)
                                    \left(\hat{a}_{\sigma}(t,t')+\hat{a}_{\sigma}^{\dagger}(t,t')\right)
                                    \hat{\rho}_{\sigma\sigma'}(t)
                                    -g_{\sigma}g_{\sigma'}\left(\hat{a}_{\sigma}(t,t')+\hat{a}_{\sigma}^{\dagger}(t,t')\right)
       \hat{\rho}_{\sigma\sigma'}(t)
       \left(\hat{a}_{\sigma'}+\hat{a}_{\sigma'}^{\dagger}\right)\right)\nonumber\\
 &-&\left(g_{\sigma}g_{\sigma'}\left(\hat{a}_{\sigma}+\hat{a}_{\sigma}^{\dagger}\right)
                                                \hat{\rho}_{\sigma\sigma'}(t)
                                                \left(\hat{a}_{\sigma'}(t,t')+\hat{a}_{\sigma'}^{\dagger}(t,t')\right)\
 -g_{\sigma'}^2\hat{\rho}_{\sigma\sigma'}(t)
                        \left(\hat{a}_{\sigma'}(t,t')+\hat{a}_{\sigma'}^{\dagger}(t,t')\right)
                        \left(\hat{a}_{\sigma'}+\hat{a}_{\sigma'}^{\dagger}\right)\right)\nonumber\\
     &&\left(\mathbbm{e}^{-\mathbbm{i}\omega (t-t')}n(\omega)+\mathbbm{e}^{\mathbbm{i}\omega (t-t')}(n(\omega)+1)\right)\!\Big\},\label{mastereq_HO}
\end{eqnarray}
\end{widetext}
 where $g_{\sigma}=\sqrt{\hbar/(2m\Omega_{\sigma})}$ and $n(\omega)$ is the Bose function. We observe that the equations of motion for the four components of $\hat{\rho}_S$ are not coupled to each other. This is the consequence of neglecting the tunneling in the qubit Hamiltonian Eq.~(\ref{sysH}).  While the two diagonal components fulfill the same equations of motion as in the case of the well-known damped harmonic oscillator, each of them with a different frequency, the two off-diagonal elements of the density matrix have a more complicated evolution. Specifically, they are not Hermitian and do not conserve the norm. This is to be expected, as the norm of the off-diagonal elements measures the qubit coherence, which is not conserved during measurement. 

One can handle master equations in an infinite-dimensional Hilbert space with
the aid of phase-space pseudoprobability distribution functions \cite{Glauber63,Cahill69,Schleich01},
which encode any operator with a finite norm \cite{Cahill69,Comment} into a phase-space function. Here, we choose the characteristic  function of the Wigner function $\chi(\alpha, \alpha^*,t)$ to represent  the density matrix
\begin{eqnarray}
\hat{\rho}_{\sigma\sigma'}(t)&=&\frac{1}{\pi}\int d^2\alpha\:\:\chi_{\sigma\sigma'}(\alpha,\alpha^*,t)\hat{D}(-\alpha)\label{wigner},
\end{eqnarray}
where $\hat{D}(-\alpha)=\exp(-\alpha \hat{A}_0^{\dagger}+\alpha^*\hat{A}_0)$ is the displacement operator. By replacing  this representation of $\hat{\rho}_{\sigma\sigma'}$ into the master equation (\ref{mastereq_HO}) we obtain partial differential equations for the characteristic functions $\chi_{\sigma\sigma'}(\alpha,\alpha^*,t)$.
Note that here $|\alpha\rangle$ is different from the 
coherent state as it is composed of Floquet modes instead of Fock states, 
i.e.\ 
\begin{eqnarray}
|\alpha\rangle&=&\mathbbm{e}^{-|\alpha|^2/2}\sum_n\frac{\alpha^n}{\sqrt{n!}}|\Phi_n(t)\rangle .
\end{eqnarray}

\section{Results}\label{section_results}

\subsection{Measurement} 
In this section we propose three dispersive measurement protocols, all based on the detection of the oscillator momentum, from which the state of the qubit can be inferred.

We start by computing the measured observable of the detector, 
the voltage drop $V$ across the SQUID. In our notation, the operator is found
as the oscillator momentum, $V=\mathbbm{i}V_0 \hat{\mathbbm{1}}_{\rm qubit}\otimes \left(a-a^\dagger\right)$, 
and involves a trace over the qubit. Here the momentum $\hat{p}$ is $2meV/\hbar$.
Thus, we obtain the diagonal characteristic functions $\chi_{\sigma\sigma}$. For $\sigma=\sigma'$  we obtain from Eqs.~(\ref{mastereq_HO}), (\ref{wigner}), and (\ref{creation0}), a Fokker-Planck equation \cite{Risken96}
\begin{eqnarray}
\dot{\chi}_{\sigma\sigma}(\alpha,\alpha^*,t)&=&
\Bigg[\left(\alpha\left(-\frac{\kappa}{2}+\mathbbm{i}\tilde{\Omega}_{\sigma}^{+}\right)+\alpha^*\left(-\frac{\kappa}{2}+\mathbbm{i}\tilde{\Omega}_{\sigma}^{-}\right)\right)\partial_{\alpha}\nonumber\\
&+&\left(\alpha^*\left(-\frac{\kappa}{2}-\mathbbm{i}\tilde{\Omega}_{\sigma}^{+}\right)+\alpha\left(-\frac{\kappa}{2}-\mathbbm{i}\tilde{\Omega}_{\sigma}^{-}\right)\right)\partial_{\alpha^*}\nonumber\\
&-&(1+2n_{\sigma})\frac{\kappa\Omega_{\sigma}}{4\Omega}(\alpha+\alpha^*)^2\nonumber\\
&+&(\alpha+\alpha^*)f_{\sigma}(t)\Bigg]\chi_{\sigma\sigma}(\alpha,\alpha^*,t),\label{chi_diag}
\end{eqnarray} 
where
\begin{eqnarray}
f_{\sigma}(t)&=&\frac{\mathbbm{i} F_0 \left(\cos (\nu t )\kappa  \nu+\sin (\nu t) \left(\Omega _{\sigma}^2-\Omega ^2\right)\right)}{\sqrt{2 m \Omega  \hbar} \left(\Omega ^2-\nu ^2\right)},
\end{eqnarray}
and $\tilde{\Omega}_{\sigma}^{\pm}=(\pm\Omega^2+\Omega_{\sigma}^2)/(2\Omega)$.  Note that we must express the operators in Eq.~(\ref{mastereq}) in terms of $\hat{A}_0$ corresponding to frequency $\Omega$ as the oscillator has initially that frequency, in a particular case the thermal state of frequency $\Omega$. Eq.~(\ref{chi_diag}) is consistent with the property $\chi_{\sigma\sigma'}(\alpha)=\chi_{\sigma'\sigma}^*(-\alpha)$ originating in the hermiticity of the density matrix. We perform the variable transformation $(\alpha,\alpha^*,t)\to(z,z^*,s)$ defined by means of following differential equations
\begin{eqnarray}
\partial_s
\alpha&=&\alpha\left(\frac{\kappa}{2}-\mathbbm{i}\tilde{\Omega}_{\sigma+}\right)
                        +\alpha^*\left(\frac{\kappa}{2}-i\tilde{\Omega}_{\sigma-}\right),\\
\partial_s\alpha^*&=&\alpha^*\left(\frac{\kappa}{2}+\mathbbm{i}\tilde{\Omega}_{\sigma+}\right)
                        +\alpha\left(\frac{\kappa}{2}+i\tilde{\Omega}_{\sigma-}\right),\\
s&=&t,
\end{eqnarray}
The solutions of these coupled differential equations will depend on some initial conditions i.e.~$\alpha(s=0)=z$ and thus we obtain the transformation $\alpha\to\alpha(z,z^*,s)$. This transformation conveniently removes the partial derivatives with respect to $\alpha$ and $\alpha^*$ in Eq.~(\ref{chi_diag}) and we are left with
\begin{eqnarray}
\partial_s  {\chi}_{\sigma\sigma} (z,z^*,s)=(\alpha(z,z^*,s)+\alpha^*(z,z^*,s))f_{\sigma}(t) \nonumber \\ \!\!\!\!
\!\!\! - (1+2n_{\sigma})\frac{\kappa\Omega_{\sigma}}{4\Omega}\left(\alpha(z,z^*,s)+\alpha^*(z,z^*,s)\right)^2,
\end{eqnarray} 
which can be directly integrated. After performing the transformation back to the initial variables $\chi_{\sigma\sigma}(\alpha,\alpha^*,t)$  we can calculate the probability density of momentum $P(p_0,t)=\sqrt{\hbar m\Omega/2}\langle\delta(\hat{p}-p_0)\rangle$ where the qubit initial state is $q_{\uparrow}|\uparrow\rangle+q_{\downarrow}|\downarrow\rangle$
and
\begin{eqnarray}
P(p_0,t)&=&\mu\!\!\!\!\sum_{\sigma\in\{\uparrow,\downarrow\}}\!\!\!\frac{|q_{\sigma}|^2}{2\pi^2}
    \int\!\!dk\:\mathbbm{e}^{\mathbbm{i}k(\dot{\xi}(t)m-p_0)}\!\!\int\!\! d^2\alpha\: \chi_{\sigma\sigma}(\alpha,\alpha^*,t)\nonumber\\
&\times&\int\!\!d^2\beta\:\langle
\beta|\mathbbm{e}^{-k\mu(\hat{A}_0^{\dagger}-\hat{A}_0)} \hat{D}(-\alpha)|\beta\rangle \label{momentum},\\
\mu&=&\sqrt{\frac{\hbar m\Omega}{2}}.
\end{eqnarray} 
Evaluating the integrals in Eq.~(\ref{momentum}) we obtain for the probability density of momentum
\begin{equation} P(p_0,t)=
\sum_{\sigma\in\{\uparrow,\downarrow\}}\frac{|q_{\sigma}|^2}{\sqrt{4 \pi B_{\sigma}(t)}}\exp\left(-\frac{(p_0-C_{\sigma}(t))^2}{4\mu^2 B_\sigma(t)}\right).\label{probdistrib}
\end{equation}
Here, assuming the oscillator initially in a thermal state, we have
\begin{eqnarray}
C_{\sigma}(t)=&&\!\!\!\!\!\frac{F_0\nu}{\kappa^2\nu^2+(\nu^2-\Omega_\sigma^2)^2}\bigg(\cos(\nu t)(\Omega_\sigma^2-\nu^2)+\sin(\nu t)\kappa \nu\nonumber\\
+
\mathbbm{e}^{-\kappa t/2}&&\!\!\!\!\!\!\!\!\cos({\overline{\Omega}_\sigma t)}\frac{\nu^2(\kappa^2+\Omega^2)-(\nu^2+\Omega^2)\Omega_\sigma^2+\Omega_\sigma^4}{\Omega^2-\nu^2}\label{eq:center}\\
-\mathbbm{e}^{-\kappa t/2}&&\!\!\!\!\!\!\!\!\sin({\overline{\Omega}_\sigma t)}\frac{\kappa(\Omega_\sigma^4+\nu^2(\kappa^2+\Omega^2)+\Omega_\sigma^2(\Omega^2-3\nu^2))}{2\overline{\Omega}_\sigma(\Omega^2-\nu^2)}\bigg),\nonumber
\end{eqnarray}
\begin{eqnarray}
B_{\sigma}(t)&=&\frac{1+2n(\Omega_{\sigma})}{2}\frac{\Omega_{\sigma}}{\Omega}
\Big(1
      -\mathbbm{e}^{-\kappa t}\frac{\Omega_{\sigma}^2}{\overline{\Omega}_{\sigma}^2}\nonumber\\
&+&\mathbbm{e}^{-\kappa t}\cos(2\overline{\Omega}_{\sigma}t)\frac{\kappa^2}{4\overline{\Omega}_{\sigma}^2}
      +\mathbbm{e}^{-\kappa t}\sin(2\overline{\Omega}_{\sigma}t)\frac{\kappa}{2\overline{\Omega}_{\sigma}}\Big)\nonumber\\
      &-&\frac{1+2n(\Omega)}{2}\mathbbm{e}^{-\kappa t}\Big( -\frac{\Omega_{\sigma}^2}{2\overline{\Omega}_{\sigma}^2}
           \left(1+\frac{\Omega_{\sigma}^2}{\Omega^2}\right)\\
      &+&\cos(2\overline{\Omega}_{\sigma}t)\frac{\Omega_\sigma^2\left(1+\frac{\Omega_\sigma^2}{\Omega^2}\right)-4\overline{\Omega}_{\sigma}^2}{4\overline{\Omega}_{\sigma}^2}
      +\sin(2\overline{\Omega}_{\sigma}t)\frac{\kappa}{2\overline{\Omega}_{\sigma}}\Big).\nonumber
\end{eqnarray}
and $\overline{\Omega}_{\sigma}=\sqrt{\Omega_\sigma^2-\kappa^2/4}$. One can see that $B(t)$ evolves from $B_{\sigma}(0)=1/2+n(\Omega)$ to 
$B_{\sigma}(\infty)=(1/2+n(\Omega_{\sigma}))\Omega_{\sigma}/\Omega$ and for $\Omega=\Omega_{\sigma}$  $\langle\hat{p}\rangle_{\sigma}(t)=C_{\sigma}(t)$ becomes the momentum of the classical damped oscillator with the initial conditions $\dot{p}(0)=-F_0 \kappa \nu/(\Omega^2-\nu^2)$ and $p(0)=F_0 \nu/(\Omega^2-\nu^2)$. Note that the value $B_{\sigma}(\infty)$ is independent of the initial $B_{\sigma}(0)$. Therefore the long time value of $B$  is the same also for ground and coherent state.

In the following, when analyzing different types of measurement protocols, we have to differentiate between {\em discrimination} and {\em measurement} time. Measurement time is the total time needed to transfer the information from the qubit to the observer. In a sample-and-hold protocol, one imprints the qubit state into the oscillator, then decouples the two and observes the latter.  The time needed for the first step is called discrimination time.

\subsubsection{Long time, single shot measurement}\label{ltss}
In the measurement scheme of Ref.~\cite{Lupascu04}  one needs the voltage amplitudes corresponding to the two qubit states. For this one must wait until the transients in the momentum (voltage) oscillations have died out.  From the amplitude of momentum one can then determine the state of the qubit.

Following Ref.~\cite{Lupascu04} we define the measurement time as the time required to obtain enough information to infer the qubit state
\begin{eqnarray}
\tau_{\rm m}&=&\frac{S_V}{(V_{\uparrow}-V_{\downarrow})^2},
\end{eqnarray}
where $V_{\sigma}$ is the amplitude of  the voltage for the qubit in the state  $|\sigma\rangle$ and $S_V=2k_B T R$ is the spectral density of the detector output. This is the time needed for discriminating two-long time amplitudes relative
to a noise backgroud given by $S_V$. Therefore, in our notation,
\begin{eqnarray}
\tau_{\rm m}&=&\frac{b}{\kappa(\mathcal{A}_{\uparrow}-\mathcal{A}_{\downarrow})^2},
\end{eqnarray}
where
\begin{eqnarray}
\mathcal{A}_{\sigma}&=&\frac{\nu}{\sqrt{\kappa^2\nu^2+(\nu^2-\Omega_{\sigma}^2)^2}}
\end{eqnarray}
and  $b=k_BTC_s/I_B^2$. Note that in this type of amplitude measurement it is advantageous to drive far from resonance, since at resonance the amplitudes $\mathcal{A}_{\sigma}$ become identical for the two qubit states. Off-resonance $\tau_{\rm m}$ is a monotonically falling function of $\Delta$, i.e.~larger coupling leads 
to faster measurement. Close to resonance $\tau_{\rm m}$ grows again for large values of $\Delta$. 

It is known that, when an harmonic oscillator is driven close to resonance, a {\em phase} measurement reveals most of the information about the oscillator frequency and
leads to the best resolution and quantum limited measurement. Along the lines of Ref.~\cite{Clerk03} one can suppose that, for measurement closest to the quantum limit, the conjugate observable to the one being measured should deliver no information. In our case, for off-resonant driving and amplitude measurement, most of the information about the qubit is contained in the amplitude and almost none in the phase.

\subsubsection{Short time, single shot measurement}\label{stss}

In the measurement protocol of the previous section and Ref.~\cite{Lupascu04} the desired information is extracted from the long-time $C_{\sigma}=\langle\hat{p}\rangle_{\sigma}(t)$. The method has the advantage of being "single shot", but disadvantages resulting from long time coupling to the environment such as dephasing, relaxation and loss of visibility \cite{Wilhelm05, Zhang06}.

\begin{figure}[!h]
  a)\includegraphics[width=.45\textwidth, height=0.25\textheight]{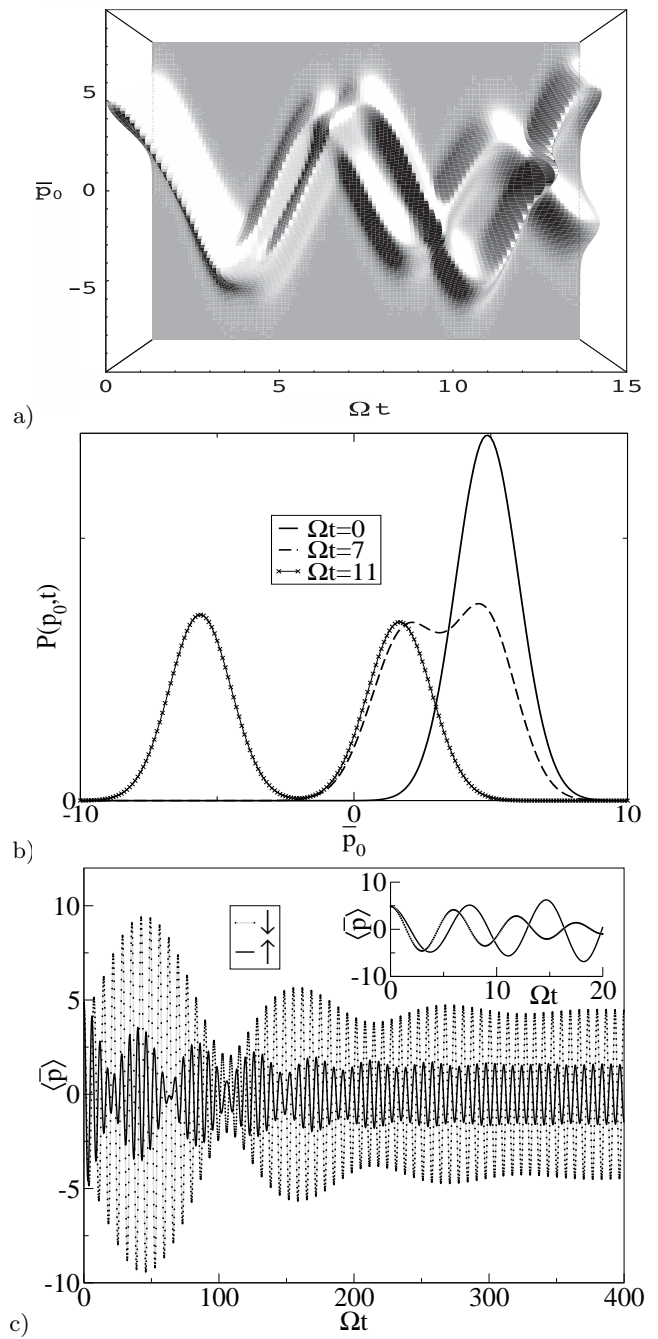} \\
    b)\includegraphics[width=.45\textwidth, height=0.24\textheight]{snapshot.eps} \\
 c) \includegraphics[width=.45\textwidth]{center.eps} 
  \caption{Probability density of momentum $P(p_0,t)$ (a), snapshots of it at different times (b) and expectation value of momentum for the two different qubit states (c). Here $\hbar\Omega/(k_BT)=2$,  $\Delta/\Omega=0.45$ , $\kappa/\Omega=0.025$ and $\hbar\nu/(k_BT)=1.9$ and $\overline{p}_0$ is the dimensionless momentum $p_0/\sqrt{k_BTm}.$}\label{probability}
\end{figure}

In this section we present a different measurement protocol. It is based
on the short time dynamics illustrated as follows:
for the qubit initially in the state $1/\sqrt{2}(|\uparrow\rangle+|\downarrow\rangle)$  the probability distribution of momentum is plotted in Fig.~\ref{probability} (a) and (b).

In Fig.~\ref{probability} one can see that the two peaks corresponding to the two states of the qubit split already during the transient motion of $\langle \hat{p}\rangle(t)$, much faster than the transient decay time. If the peaks are well enough separated, a single measurement of momentum gives the needed information about the qubit state, and has the advantage of avoiding decoeherence effects resulting from a long time coupling to the environment. 
Nevertheless the parameters we need to reduce the discrimination time also enhance the decoherence rate. 

We define in this case the discrimination time as the first time when the two peaks are separated by more than the sum of their widths i.e.
\begin{eqnarray}
|C_{\uparrow}(\tau_{\rm discr})-C_{\downarrow}(\tau_{\rm discr})|\nonumber\\
\geq3 \sqrt{2 m\hbar\Omega} \left(\sqrt{B_{\uparrow}(\tau_{\rm discr}})+\sqrt{B_{\downarrow}(\tau_{\rm discr})}\right)\label{eq:tau}.
\end{eqnarray}
A comparison between discrimination and dephasing rate will be given in Section \ref{comparison_section}.

Because of the oscillatory nature of  $C_{\sigma}(t)$ the problem of finding the first root of Eq.~(\ref{eq:tau}) is not trivial. We solve it by semi-quantitatively probing the function $|C_{\uparrow}(t)-C_{\downarrow}(t)|-3\sqrt{2 m\hbar\Omega}(\sqrt{B_{\uparrow}(t)}+\sqrt{B_{\downarrow}(t)})$ therefore the plot ist not very accurate. Nevertheless it gives a good idea about the dependence of $\tau_{\rm discr}$ on $\Delta$. 
 \begin{figure}[!h]
  \includegraphics[width=.4\textwidth]{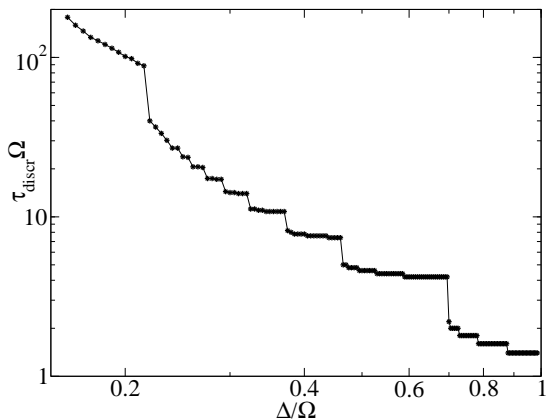} 
  \caption{Discrimination time as function of the coupling strength between qubit and oscillator. Here $\hbar\Omega/(k_BT)=2$, $\kappa/\Omega=0.025$ and $\hbar\nu/(k_BT)=1.95$.}\label{tdiscr1}
\end{figure}

We observe in Fig.~\ref {tdiscr1} that $\tau_{\rm discr}$ is a discontinuous function of the coupling strength $\Delta$, such that small adjustments in the parameters can give important improvement of the discrimination time. 

For this type of measurement we are interested in the transients of $C_{\sigma}(t)$ and we observe that the difference $|C_{\uparrow}(t)-C_{\downarrow}(t)|$ increases for values of the driving frequency $\nu$ close to resonance. For the $\nu$ far off-resonance the splitting of the peaks is increased by stronger driving. 

The discrimination time discussed here is not to be confused
with the physical measurement time. In particular, the discrimination time
remains finite even at vanishing $\kappa$ and when the
off-diagonal elements of the full density matrix in qubit space (\ref{densityqubit}) still have finite norm at this time. The discrimination time is the time it takes to imprint the 
qubit state into the oscillator dynamics. For completing the measurement 
the oscillator itself needs to be observed by the heat bath and, as a consequence of that observation, the full density matrix will collapse further. 

 We note that in this kind of sample-and-hold measurement, the qubit spends only the discrimination time in contact with the environment. Keeping the discrimination time short may be of advantage in limiting bit flip errors during detection. We do not further describe such error processes in this paper.
 
As a technical limitation, it should be remembered that our theory is based on a
Markov approximation for the oscillator-bath coupling, hence it is not reliable
for discrimination times lower than the bath correlation time. 

\subsubsection{Quasi-instantaneous, ensemble  measurement}\label{instantan}

In this section, we are going to take this idea to the next level and analyze 
a measurement protocol that is based on extremely short qubit-detector 
interaction. 
In Refs.~\cite{Lougovski06,Bastin06b,FrancaSantos07} has been shown that one can measure several field observables through infinitesimal-time probing of the internal states of the coupled qubit. In this section, we apply the same idea to the opposite setting. We show that the information about the state of the qubit is encoded in the expectation value of the momentum of our oscillator at only one point in time, leading to a fast weak measurement scheme.

We rewrite the Hamiltonian (\ref{bathH})
\begin{eqnarray}
\hat{H}&=&\hbar\Omega\left(\hat{a}^{\dagger}\hat{a}+\frac{1}{2}\right)
               + F(t)\sqrt{\frac{\hbar}{2m\Omega}}\left(\hat{a}+\hat{a}^{\dagger}\right)
               + \epsilon(t)\hat{\sigma}_z\nonumber\\
               &+& \hat{\sigma}_z\frac{\hbar \Delta^2}{4\Omega}\left(\hat{a}+\hat{a}^{\dagger}\right)^2
               + \frac{\hbar\left(\hat{a}+\hat{a}^{\dagger}\right)}{2\sqrt{m\Omega}}                  \sum_i\frac{\lambda_i\left(\hat{b}_i+\hat{b}_i^{\dagger}\right)}{\sqrt{m_i\omega_i}}\nonumber\\
               &+&\sum_i\hbar\omega_i\left(\hat{b}_i^{\dagger}\hat{b}_i+\frac{1}{2}\right).
\end{eqnarray}
In Schr\"odinger picture we have
\begin{eqnarray}
\dot{\hat{\rho}}&=&\frac{1}{\mathbbm{i}\hbar}[\hat{H},\hat{\rho}],
\end{eqnarray}
which leads, for any observable, to
\begin{eqnarray}
\partial_t\langle\hat{O}\rangle&=&\langle \partial_t\hat{O}\rangle+\frac{1}{\mathbbm{i}\hbar}\langle[\hat{O},\hat{H}]\rangle.
\end{eqnarray}
Setting $\hat{O}=\hat{a}-\hat{a}^{\dagger}$ we obtain:
\begin{eqnarray}
\partial_t\langle \hat{a}-\hat{a}^{\dagger}\rangle&=&\frac{1}{\mathbbm{i}\hbar}\Bigg\langle \hbar \Omega(\hat{a}+\hat{a}^{\dagger})
+\frac{\hbar\Delta^2}{2\Omega}\hat{\sigma}_z(\hat{a}+\hat{a}^{\dagger})\label{derivative}\\
 &-& 2\sqrt{\frac{\hbar}{2m\Omega}}F(t)
 -2\sqrt{\frac{\hbar}{2m\omega_i}}\lambda_i\left(\hat{b}_i+\hat{b}_i^{\dagger}\right)\Bigg\rangle\nonumber.
\end{eqnarray}
We assume unbiased noise and the qubit in the pure initial state $q_{\uparrow}|\uparrow\rangle+ q_{\downarrow}|\downarrow\rangle$ which leads to 
\begin{eqnarray}
\hat{\rho}_{\rm qubit}(0)=\left(\begin{matrix}|q_{\uparrow}|^2 & q_{\uparrow} q_{\downarrow}^*\\
q_{\uparrow}^* q_{\downarrow} &|q_{\downarrow}|^2\end{matrix}\right).
\end{eqnarray}
For $t=0$ Eq.~(\ref{derivative}) becomes
\begin{eqnarray}
\partial_t\langle \hat{a}-\hat{a}^{\dagger}\rangle|_{t=0}
  &=&-\mathbbm{i}\left \langle \hat{a} + \hat{a}^{\dagger}\right \rangle_{t = 0}
          \left(\Omega+\frac{\Delta^2}{2\Omega}\langle\hat{\sigma}_z\rangle\right)\nonumber\\
          &-&2\sqrt{\frac{\hbar}{2m\Omega}}F(0)\\
\langle \hat{\sigma}_z  \rangle&=&2|q_{\uparrow}|^2-1       
  \end{eqnarray}
If $\left \langle \hat{a} + \hat{a}^{\dagger}\right \rangle_{t = 0}\not =0$, the ensemble measurement of $\partial_t\langle \hat{a}-\hat{a}^{\dagger}\rangle|_{t=0}$ is sufficient to determine the state of the qubit. 

In our case the oscillator is initially in a thermal state and $\left \langle \hat{a} + \hat{a}^{\dagger}\right \rangle_{t = 0}=0$. Nevertheless, calculating $\langle \hat{p} \rangle$ from Eq.~(\ref{probdistrib}) we obtain for the center of the Gaussians corresponding to the two qubit states $C_{\sigma}$ of Eq.~(\ref{eq:center})
\begin{eqnarray}
\langle p \rangle(t) &=& C_{\downarrow}(t)+(C_{\uparrow}(t)-C_{\downarrow}(t))|q_{\uparrow}|^2,
\end{eqnarray}
which is valid for {\em all} times. If $C_{\uparrow}(t)\not=C_{\downarrow}(t)$ we have
\begin{eqnarray}
|q_{\uparrow}|^2&=&\frac{\langle p \rangle(t)-C_{\downarrow}(t)}{C_{\uparrow}(t)-C_{\downarrow}(t)}.\label{qubitstate}
\end{eqnarray}
At $t=0$ we have like in the exact case $C_{\uparrow}(0)=C_{\downarrow}(0)$ independent of system parameters. This is again the consequence of the thermal initial state. Therefore one cannot infer from $\langle \hat{p} \rangle(0)$ the state of the qubit. For infinitesimal $\tau>0$ we have $C_{\uparrow}(\tau)\not =C_{\downarrow}(\tau)$. Therefore quasi-instantaneous measurement of momentum still delivers the necessary information about the qubit, if the measurement is made at a infinitesimally small $\tau>0$. 

At $t=0$ also the first derivative of $C_{\uparrow}(t)-C_{\downarrow}(t)$ is 0 due to the thermal initial state. A series expansion of Eq.~(\ref{eq:center}) gives the
short time result 
\begin{eqnarray}
C_{\uparrow}(t)-C_{\downarrow}(t)&=&\frac{2\nu F_0\Delta^2}{\nu^2-\Omega^2}\cdot t^2+\mathcal{O}(t^3)\label{createinitial}.
\end{eqnarray}
This gives a criterion for $\tau_{discr}$, independent of $\kappa$, similar to observations of previous section, i.e. for short discrimination times we need large $\Delta$ and strong, close to resonance driving. 

Moreover, it is sufficient to measure the expectation value of momentum, and not the first time derivative. 
The reason for this is the oscillator evolution, mediated by the interaction with the qubit, into a state with finite expectation value $\left\langle a+a^\dagger \right\rangle$, in other words the system is automatically creating its own measurement favorable "initial" condition. This is visible in Eq.~(\ref{createinitial}) where the part of the signal proportional $|q_{\uparrow}|^2$ increases like $t^2$ which, for short times is slower than $t$, as it would be in the case where the favorable initial condition already exists.

This method leads to shorter discrimination times than the protocols presented in section \ref{ltss} and \ref{stss} which are independent of $\Delta$ and $\nu$.
Again, the read out of the oscillator in the end will be a separate issue and
ultimately take a time $\propto \kappa^{-1}$.  

On the other hand, in order to establish the expectation value with sufficient
accuracy, this scheme requires a large ensemble average. According to the central limit theorem the uncertainty of the ensemble measurement is 
\begin{eqnarray}
\frac{\Delta y}{\langle y\rangle}&=&\frac{1}{\sqrt{N}}\frac{\Delta p}{\langle p\rangle}
\end{eqnarray}
where $y$ is the ensemble averaged value of the measured momentum and $N$ the number of measurements. For a given precision we have $N\propto\left(\Delta p/\langle p\rangle\right)^2$, therefore the number of measurements necessary to reach a given precision depends on $\Delta$ and time $t$. We have
\begin{eqnarray}
N=\frac{\langle y\rangle^2}{\Delta y^2}\frac{|q_{\uparrow}|^2|q_\downarrow|^2(C_{\uparrow}-C_{\downarrow})^2+m\Omega\hbar\sum_{\sigma\in\{\uparrow,\downarrow\}}|q_{\sigma}|^2B_{\sigma}}{\left(\sum_{\sigma\in\{\uparrow,\downarrow\}}|q_{\sigma}|^2C_{\sigma}(t)\right)^2}
\end{eqnarray}
Eq. (\ref{qubitstate}) shows that we need $C_{\uparrow}\not=C_{\downarrow}$ in order to determine the state of the qubit. At the same time, the number of measurements $N$ necessary for high precision measurement of momentum is significantly reduced when $C_{\uparrow}=C_{\downarrow}$ (the two Gaussian distributions overlap completely). This reflects the tradeoff between the number of measurements and the signal strength $C_{\uparrow}-C_{\downarrow}$ which provides the information about the qubit.

\subsection{Back-action on the qubit}

In order to complete the study  of the measurement protocols presented in the previous section, we need insight into the measurement bakaction on the qubit.  Since we are studying the QND Hamiltonian (\ref{sysH}), the qubit decoherence consists only of dephasing. We start with the qubit in the initial pure state $( |\uparrow\rangle+|\downarrow\rangle ) /\sqrt{2}$ and study the decay of the off-diagonal elements of the qubit density matrix.  Such a superposition can be created by e.g.~rapidly switching the tunnel matrix element from a large value to zero \cite{Koch06}, or by ramping up the energy bias from zero to a large value. We compute the qubit coherence
\begin{eqnarray} 
C(t) & = & {\rm Tr} \left(\hat{\sigma}_x\otimes \hat{\mathbbm{1}}\hat{\rho}_S(t)\right)  
= 2 {\rm Re}{\rm Tr}\hat{\rho}_{\uparrow\downarrow}(t)\nonumber\\
          & = & 2 {\rm Re}\int\! dx\langle x|\hat{\rho}_{\uparrow\downarrow}(t)|x\rangle\nonumber\\
          & = & 2 {\rm Re}\int\! dx\!\int\! dp\:W_{\uparrow\downarrow}(x,p,t)\nonumber\\ 
          & = &2 {\rm Re}\int\! dx\!\int\! dp\:e^{\mathbbm{i}x0}e^{\mathbbm{i}p0}W_{\uparrow\downarrow}(x,p,t)\nonumber\\
         &=& 8\pi {\rm Re}\:\chi_{\uparrow\downarrow}(0,0,t),\label{coherence}
\end{eqnarray} 
where $W_{\sigma\sigma'}$ is the Wigner function  
\begin{eqnarray} 
\!\!\!\! W_{\sigma\sigma'}(x_0,p_0)&=&\frac{1}{\pi\hbar}\int dy\langle x_0+y|\hat{\rho}_{\sigma\sigma'}e^{-2\mathbbm{i}yp_0}|x_0-y\rangle . \nonumber \\
\end{eqnarray} 
We extract the dephasing time  $\tau_\phi$ from the strictly exponential
long-time tail of $C(t)$.

We rewrite the master equation (\ref{mastereq_HO}) for $\sigma\not =\sigma'$ using Eqs.~(\ref{creation0}) and (\ref{wigner}) and obtain a partial differential equation for the characteristic function $\chi_{\uparrow\downarrow}$
\begin{eqnarray}
\dot{\chi}_{\uparrow\downarrow}(\alpha,\alpha^*,t) & = & 
    \bigg(\left(\alpha(k_1+\mathbbm{i}\Omega)+\alpha^*k_1+ B\sin(\nu t)\right)\partial_{\alpha}\nonumber\\
&+& \left(\alpha^*(k_2-\mathbbm{i}\Omega)+\alpha k_2-B\sin(\nu t)\right)\partial_{\alpha^*}\nonumber\\
&-& i\frac{\Delta^2}{2\Omega}(\partial_{\alpha}-\partial_{\alpha^*})^2
+  ({\alpha}+{\alpha^*})f_{\uparrow\downarrow}(t)+\mathcal{F}(t) \nonumber\\
&+& p (\alpha+\alpha^*)^2\bigg)\chi_{\uparrow\downarrow}(\alpha,\alpha^*,t)\label{offdiag}.
\end{eqnarray} 
The coefficients  can be found in Appendix \ref{ApFP}. Eq.~(\ref{offdiag}) is a generalized Fokker-Planck equation where the total norm is not conserved, i.e.~$\int\! d^2\alpha \chi_{\uparrow\downarrow}(\alpha,\alpha^{\ast},t)$ is {\em not} a constant of  motion.  

\subsubsection{Solution of the generalized Fokker-Planck equation}

Generalized Fokker-Planck equations (\ref{offdiag}) cannot in general
be solved analytically with the established tools \cite{Risken96}. 
In our case, we are not interested in a fully general solution of the differential equation, but in the initial value problem where the $\chi_{\uparrow\downarrow}(\alpha,\alpha^\ast,0)$ is a Gaussian function, which covers thermal and coherent states of the oscillator. In this case one can show that $\chi_{\uparrow\downarrow}(\alpha,\alpha^\ast,t)$ remains a Gaussian at all time. This is the consequence of the QND Hamiltonian (\ref{sysH}). We make the ansatz 
\begin{eqnarray}
\chi_{\uparrow\downarrow}(\alpha,\alpha^*,t) & = & A(t)\exp\big(-M(t)
\alpha^2-N(t)\alpha^{*2}\nonumber\\
&-&Q(t)\alpha\alpha^*+R(t)\alpha+S(t)\alpha^*\big),\label{ansatz}
\end{eqnarray} 
and obtain for the time-dependent  parameters of the Gaussian a closed system of nonlinear differential equations of the first order, thus proving that our ansatz is correct and complete if the initial characteristic function is a Gaussian.  

We assume the oscillator initialy in a thermal state $\chi_{\uparrow\downarrow}(\alpha,\alpha^*,0)=(1/4\pi)\exp(-(1/2+n(\Omega))|\alpha|^2)$ and obtain for the parameters of the Gaussian ansatz  following equations of motion
\begin{eqnarray}
\dot{A}_E(t)&=&B\sin(\nu t)(R(t)-S(t))+\mathcal{F}(t)\nonumber\\
&+&\frac{\mathbbm{i}\Delta^2}{\Omega}\left(M(t)+N(t)-Q(t)-\frac{(R(t)-S(t))^2}{2}\right),\label{eq1}\\
\dot{R}(t)&=&(k_1+\mathbbm{i}\Omega)R(t)+k_2S(t)-B\sin(\nu t)(2M(t)-Q(t))\nonumber\\
&-&\frac{\mathbbm{i}\Delta^2}{\Omega}(R(t)-S(t))(Q(t)-2M(t))+f_{\uparrow\downarrow}(t)\\
\dot{S}(t)&=&(k_2-\mathbbm{i}\Omega)S(t)+k_1R(t)+B\sin(\nu t)(2N(t)-Q(t))\nonumber\\
&-&\frac{\mathbbm{i}\Delta^2}{\Omega}(S(t)-R(t))(Q(t)-2~N(t))+f_{\uparrow\downarrow}(t)\\
\dot{M}(t)&=&2(k_1+\mathbbm{i}\Omega)M(t)
                     +k_2Q(t)\nonumber\\
                     &+&\frac{\mathbbm{i}\Delta^2}{2\Omega}(Q(t)-2M(t))^2
                     -p,\label{eq2}\\
\dot{N}(t)&=&2(k_2-\mathbbm{i}\Omega)N(t)
                     +k_1Q(t)\nonumber\\
                     &+&\frac{\mathbbm{i}\Delta^2}{2\Omega}(Q(t)-2N(t))^2
                     -p,\label{eq3}\\
\dot{Q}(t)&=&(k_1+k_2)Q(t)
                     +2k_1M(t)
                     +2k_2N(t)\nonumber\\
                     &-&\frac{\mathbbm{i}\Delta^2}{\Omega}(Q(t)-2M(t))(Q(t)-2N(t))-2p\label{eq4},
\end{eqnarray} 
where $A(t)=\mathbbm{e}^{A_E(t)}$. This system of equations can be solved numerically, for example using a Runge-Kutta algorithm. 

Ref.~\cite{Serban06} gives a elaborate analysis of the various dephasing mechanisms in the case without driving and the parameter regimes where they come to play. There the weak qubit-oscillator coupling (WQOC) regime is associated to a phase Purcell effect~\cite{Purcell46} where the dephasing rate $1/\tau_\phi\propto 1/\kappa$. Beyond the weak coupling, Ref.~\cite{Serban06} explores a strong dispersive coupling regime with fundamentally different origin where the dephasing rate is proportional to $\kappa$. 

In the following we want to apply and extend this results to the case of actual measurement, i.e.~when the oscillator is driven in order to measure its frequency and from this information, to infere the state of the qubit.

\subsubsection{Qubit dephasing}

We start by studying the dependence of the qubit dephasing on the parameters of the oscillator driving field. 

\begin{figure}[!h]
  \includegraphics[width=.45\textwidth]{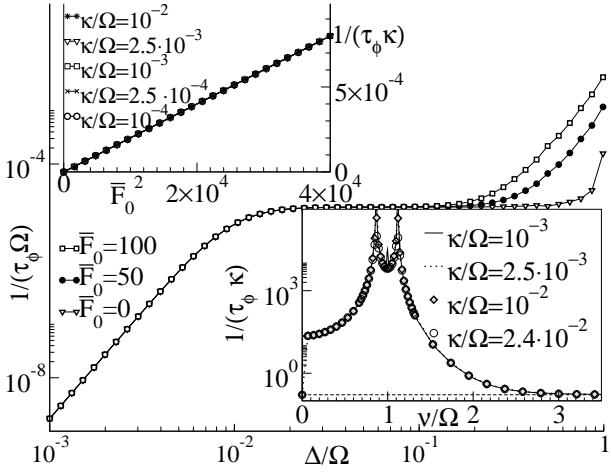} \\
   \caption{Dephasing rate dependence on driving: dependence on $\Delta$ for different driving strengths $F_0$ ($\kappa/\Omega=10^{-4}$ and $\nu=2\Omega$). Top inset: dependence of the decoherence rate on $F_0$ for different values of $\kappa$ ($\Delta/\Omega=5\cdot10^{-2}$ and $\nu=2\Omega$). Bottom inset: dependence of the decoherence rate on driving frequency $\nu$ for different vales of $\kappa$ ($\Delta/\Omega=0.5$). Here $\hbar\Omega/k_BT=2$ and $\overline{F}_0$ is the dimensionless force $F_0\hbar/(k_BT\sqrt{m k_BT})$.}  \label{antrieb}
\end{figure}

In Fig.~\ref{antrieb} we observe that the dependence of the dephasing rate $1/\tau_\phi$ on $F_0$ is quadratic. For values of $\kappa$ belonging to strong and weak coupling regime at $F_0=0$  we obtain the same driving contribution to the dephasing rate, proportional to $\kappa F_0^2$, see the inset of Fig.~\ref{antrieb}. Here only the contribution of driving is shown. We have substracted from each curve the initial value of $\tau_\phi$ at $F_0=0$. We observe that the decoherence rate must be of the form
\begin{eqnarray}
\frac{1}{\tau_{\phi}}=\frac{1}{\tau_{\phi}}\bigg |_{F_0=0}+{\rm ct.}\cdot F_0^2\kappa,
\end{eqnarray}
for both the weak and strong coupling regime. 

This was to be expected since the qubit couples to the squared coordinate which (at least in the classical case) is proportional to $F_0^2$. In both regimes, the driving leads to a contribution to the dephasing rate that is proportional to $\kappa$ because the driving leads to classical 
motion relative to the heat bath, which
is fixed in the $\hat{x}$-coordinate space. This motion enhances the effect of the 
bath the stronger the friction coefficient $\kappa$ is. Consequently,
even if in the undriven case the dephasing rate scales as $1/\kappa$, 
strong driving can in principle cross it over to a decay rate $\propto \kappa$. 
 This cross-over from $1/\kappa$ to $\kappa$ inside the WQOC regime happens at either {\em very} strong driving or when the driving $\nu$ frequency approaches one of the system resonances $\Omega_{\sigma}$.
 
The dependence on the driving frequency has also been analyzed in Fig.~\ref{antrieb}. Here we observe two peaks at $\Omega_{\uparrow}$ and $\Omega_{\downarrow}$. At $\nu=\Omega$ the classical driven and {\em un}damped trajectory $\xi(t)$ diverges. In terms of the calculation this means that the Floquet modes are not
well-defined when the driving frequency is at resonance with the harmonic oscillator --- we have a continuum instead. Physically this means that at $t=0$ our oscillator has the frequency $\Omega$ because it has not yet "seen" the qubit, and we are driving it at resonance, and by amplifying the oscillations of $\langle \hat{x}\rangle$ which is subject to noise we amplify the noise seen by the qubit. The dephasing rate is also expected to diverge. The peaks at $\Omega_{\uparrow}$and $\Omega_{\downarrow}$  show the same effect after the qubit and the oscillator become entangled. The dephasing rate drops again for large driving frequencies
to the value obtained in the case without driving. 

\subsection{Comparison of dephasing and measurement times}\label{comparison_section}

In this section we analyze the measurement times necessary for the measurement protocols described in sections \ref{ltss} and \ref{stss} and compare them with the dephasing times of the qubit obtained for the same parameters.

\subsubsection{Long time, single shot measurement}

For the long time measurement protocol (section \ref{ltss}) we observe that $1/\tau_{\rm m}\propto F_0^2\Delta^4+\mathcal{O}(\Delta^8)$. 

\begin{figure}[!h]
\includegraphics[width=.4\textwidth]{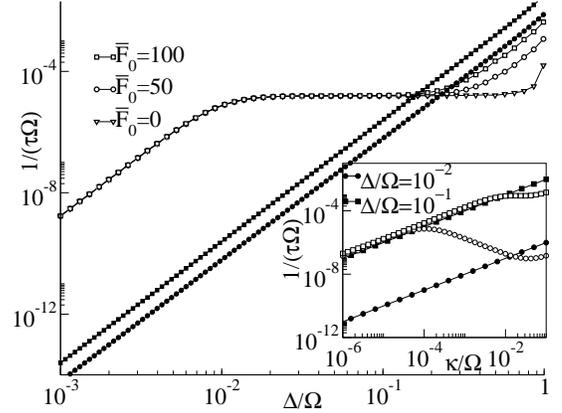}
\caption{Comparison of dephasing / measurement rate. Open symbols: $1/(\tau_\phi\Omega)$ and filled symbols: $1/(\tau_{\rm m}\Omega)$. Here $\nu=2\Omega$, $\kappa=10^{-4}\Omega$ and $\hbar\Omega/(k_B T)=2$. Inset: dependence on $\kappa$, $\overline{F}_0=200$. The crossings, conflicting with the quantum limit \cite{Braginsky95}, are signaling the limits of the Born approximation, as described in text.}\label{comp}
\end{figure}

Comparing  $1/\tau_\phi$ and $1/\tau_{\rm m}$ we find that the measurement time depends more stronlgy on the driving strength $F_0$ than the dephasing time. 

As one can see for the parameters of Fig.~\ref{comp}, in the WQOC regime the measurement time is longer than the dephasig time. Their difference decreases as we increase $\Delta$ due to the onset of the strong coupling plateau in the dephasing rate,  approaching the quantum limit where the measurement time becomes comparable to the dephasing time. Note that, for superstrong coupling either between qubit and oscillator or between oscillator and bath, corrections of the order $(\kappa/\Omega_{\downarrow})^2$ of the dephasing rate gain importance. These corrections are not treated in our Born approximation. Therefore the regions where the dephasing rate becomes lower than the measurement rate, in violation with the quantum limitation of Ref.~\cite{Braginsky95}, should be regarded as a limitation of our approximation.

The inset in Fig.~\ref{comp} shows the dephasing and measurement times as function of $\kappa$. Again we observe improvement of the ratio of measurement and dephasing time as we increase $\Delta$. On the other hand, if the tunning of $\kappa$ should be easier to achieve experimentally, we also see that, at given $\Delta$ one can make use of the phase Purcell effect, which reduces the dephasing rate as $1/\kappa$ while the measurement rate increases like $\kappa$. This goes along the lines of  Ref.~\cite{Serban06} where it has been shown that strong $\kappa$ implies WQOC, i.e. phase Purcell effect. 

 \subsubsection{Short time, single shot measurement}

\begin{figure}[!h]
\includegraphics[width=.4\textwidth]{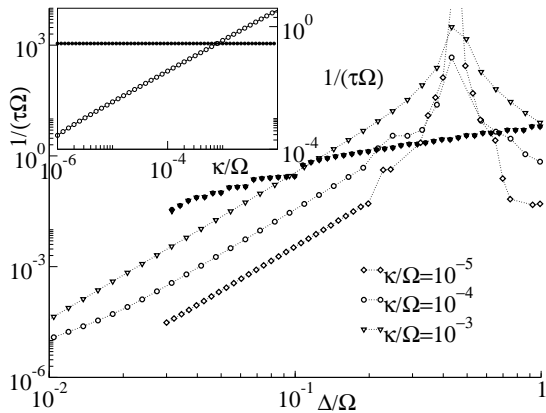}
\caption{Comparison of dephasing / discrimination rates. Open symbols: $1/(\tau_\phi\Omega)$ and filled symbols: $1/(\tau_{\rm discr}\Omega)$. Here $\nu=1.1\Omega$ ,$\overline{F}_0=200$ and $\hbar\Omega/(k_B T)=2$. Inset: dependence on $\kappa$ for $\Delta/\Omega=0.1$.}\label{shorttime}
\end{figure}

As already mentioned, for the short time, single shot measurement strong, close to resonance driving is needed for the rapid separation of the peaks. While the discrimination time is not very sensitive to the change of $\kappa$, we observe in Fig.~\ref{shorttime} that one needs relatively strong coupling ($\Delta/\Omega\in(0.03,0.1)$) for the discrimination time to become shorter than the decoherence time. The picture  of the dephasing rate is also qualitatively different from the case without driving or with far off-resonant driving, since for $\Delta/\Omega\approx 0.4$,  $\Omega_{\uparrow}$ becomes resonant with the driving frequency. In this region our numerical calculation also becomes unstable. Nevertheless, as one can see in Fig.~\ref{shorttime}, the dephasing rate is proportional to $\kappa$. Thus, by reducing the damping of the oscillator, one can extend the domain of values of $\Delta$ where the measurement can be performed. 
We also observed that by further reducing $\Delta$ the discrimination rate suddenly drops to zero, i.e. for too small $\Delta$ the two peaks in Fig.~\ref{probability} will never be well enough separated to allow a single shot measurement.

In this protocol what we call "discrimination time" is actually the time when the system becomes measurable, i.e. one can in principle extract from a single measurement the needed information about the qubit (and consequently collapse the wave function). We do not further describe this collapse here. 
\section{Conclusion}
We have presented a phase space theory of the measurement and measurement backaction on a qubit coupled to a dispersive detector.

We have studied the qubit coupled to an complex environment (weakly damped harmonic oscillator) with a quadratic coupling, which does not have to be weak.  We solved the problem by considering the prominent degree of freedom of the environment, i.e. the main oscillator as part of the quantum mechanical system and explicitly solving its dynamics. Only at the end of the calculation we traced over this last degree of freedom of the environment in order to obtain the qubit dynamics.

We presented three measurement protocols and compared the measurement and decoherence times. The protocol of section \ref{ltss} requires long measurement time, such that the measurement can only be preformed in strong coupling regime, with far off-resonant driving. The protocol of section \ref{stss} has the advantage of short  discrimination times compared with the dephasing time, requires strong qubit-oscillator coupling and also close to resonance driving. Both this protocols can be performed as single shot measurement, and thus may be useful as a readout method for the scalable architecture with long range coupling using superconducting flux qubits \cite{Fowler07}. The quasi-instantaneous measurement protocol of section \ref{instantan} has the advantage of the shortest possible discrimination time and no restriction for the qubit-oscillator coupling, with the drawback that one needs to repeat the measurement a large number of times to obtain the momentum expectation value.

We expect our results, with minor adaptations, to
be applicable to various cavity systems, e.g. quantum dot or atom-based
quantum optical schemes \cite{Raimond01,Balodato05}. The dispersive coupling of Hamiltonian (\ref{sysH}) could have implications for the generation of squeezed states, quantum memory in the frame of quantum information processing, measurement and post-selection of the number states of the cavity. 

\section{Acknowledgment}
We are grateful to F.~Marquardt  and J.~v.~Delft for very helpful discussions.  This work is supported by NSERC through a Discovery Grant and the DFG through SFB 631 and by EU through EuroSQIP and RESQ projects. I.S.~acknowledges support through the Elitenetzwerk Bayern. 

\section{Appendix}\label{Ap1}
\subsection{Parameter conversion}\label{Apcirc}
Here we present the parameter conversion from the actual circuit to our model Hamiltonian.
\begin{table}[!h]
\begin{tabular}{|c|c|c|}
\hline
Description   & Symbol   & Circuit \\ \hline
Frequency      & $\Omega$               & $\sqrt{\frac{f_1}{L_{JS}C_S}}$ \\\hline
Mass               &  m                            & $\frac{\hbar^2C_S}{2e^2}$     \\\hline
Qubit coupling & $\Delta$                  & $\sqrt{\frac{\delta f_1}{L_{JS}C_S}}$ \\\hline
Driving field     & $F(t)$                      & $I_B(t)\frac{\hbar}{e}$ \\\hline
Qubit driving    & $\epsilon(t)$           & $\epsilon_0+\upsilon I_B^2(t)$ \\\hline
Momentum      & $P_+$                     &$\frac{\hbar^2C_S}{2e^2}\dot\Gamma_+=\frac{C_S\hbar V}{e}$\\\hline
Position           &$x$                              &$\Gamma_+$  \\\hline          
Damping constant&$\kappa$&$1/(2RC_S)$\\\hline     
\end{tabular}
\caption{Parameter conversion}
\end{table}
From the relation $f(\hat{\gamma})=f(\gamma_0\hat{\sigma}_z)=\left(f(\gamma_0)+f(-\gamma_0)\right)/2+\hat{\sigma}_z\left(f(\gamma_0)-f(-\gamma_0)\right)/2$ we have:
\begin{eqnarray*}
f_1&=&\frac{1}{2}\left(\cos\left(\frac{a+b\gamma_0}{2}\right)+\cos\left(\frac{a-b\gamma_0}{2}\right)\right)\\
\delta f_1&=&\frac{1}{2}\left(\cos\left(\frac{a+b\gamma_0}{2}\right)-\cos\left(\frac{a-b\gamma_0}{2}\right)\right)\\
a&=&\Xi_1\frac{M_{\Sigma}^2}{2L_q}=\frac{2\pi}{\Phi_0}\left(-M_{Sq}\Phi_{q}^{(x)}+L_q\Phi_{S}^{(x)}\right)\frac{1}{2L_q}\\
b&=&\frac{M_{Sq}}{2L_q}\\
\upsilon&=&\frac{1}{4L_{JS}I^2_{cS}}\delta f_2\\
\delta f_2&=&\frac{1}{2}\left(\left(\cos\left(\frac{a+b\gamma_0}{2}\right)\right)^{-1}-\left(\cos\left(\frac{a-b\gamma_0}{2}\right)\right)^{-1}\right)
\end{eqnarray*}
\subsection{Parameters for the generalized Fokker-Planck equation}\label{ApFP}
\begin{eqnarray*}
k_{1,2}&=&-\frac{\kappa}{2}\pm\kappa\frac{(1+2n_{\uparrow})\Omega_{\uparrow}-(1+2n_{\downarrow})\Omega_{\downarrow}}{4\Omega}\\
p&=&-\frac{\kappa}{8\Omega}\left(\Omega_{\uparrow}(1+2n_{\uparrow})+\Omega_{\downarrow}(1+2n_{\downarrow})\right)-\frac{\mathbbm{i}\Delta^2}{8\Omega}\\
B&=&-\frac{2 \mathbbm{i} F_0\Delta^2}{\sqrt{2 m\Omega \hbar}(\Omega^2-\nu^2)}
\end{eqnarray*}
\begin{eqnarray*}
\mathcal{F}(t)&=&\frac{\mathbbm{i} F_0^2\Delta^2}{2m\hbar(\nu^2-\Omega^2)^2}\left(\cos(2\nu t)-1\right)-2\frac{\mathbbm{i}}{\hbar}\epsilon(t)\\
f_{\uparrow\downarrow}(t)&=&\frac{\kappa F_0}{\sqrt{2 m\hbar\Omega}}
\Bigg(\sin(\nu t)\Bigg(
\frac{\Delta^2}{2(\nu^2-\Omega^2)}\sum_{\sigma\in\{\uparrow,\downarrow\}}\frac{\Omega_{\sigma}(1+2n_\sigma)}{\nu^2-\Omega_\sigma^2}\nonumber\\
&-&\frac{\nu\Delta^2(1+2n_\nu)}{(\nu^2-\Omega_{\uparrow}^2)(\nu^2-\Omega_{\downarrow}^2)}\Bigg)+\frac{\mathbbm{i}\nu}{\Omega^2-\nu^2}\cos(\nu t)\Bigg)
\end{eqnarray*}

\end{document}